\documentclass[ 
11pt,%
tightenlines,%
twoside,%
onecolumn,%
nofloats,%
nobibnotes,%
nofootinbib,%
superscriptaddress,%
noshowpacs,%
centertags]%
{revtex4}

\usepackage{amssymb,amsfonts,amsmath,mathtext,enumerate,float}
\usepackage{braket}

\newtheorem{theorem}{Theorem}

\newcommand{\Tr}{{\rm Tr}\,}
\newcommand{\calH}{{\cal H}}

\begin{document}

	\title{On the construction of a quantum channel corresponding to  non-commutative graph for a qubit interacting with quantum oscillator}
	
	\author{\firstname {G.G.}~\surname {Amosov}}
	\email[E-mail:]{gramos@mi-ras.ru}
	\affiliation {Steklov Mathematical Institute of Russian Academy of Sciences, Gubkina str., 8, Moscow 119991, Russia}
	
	\author{\firstname{A.S.}~\surname {Mokeev}}
	\email[E-mail:]{aleksandrmokeev@yandex.ru}
	\affiliation {Steklov Mathematical Institute of Russian Academy of Sciences, Gubkina str., 8, Moscow 119991, Russia}
	
	\author{\firstname{A.N.}~\surname {Pechen}}
	\email[E-mail:]{apechen@gmail.com}
	\affiliation {Steklov Mathematical Institute of Russian Academy of Sciences, Gubkina str., 8, Moscow 119991, Russia}

	
	\begin{abstract}{We consider error correction, based on the theory of non-commutative graphs, for a model of a qubit interacting with quantum oscillator. The dynamics of the composite system is governed by the Schr\"odinger equation which generates positive operator-valued measure (POVM) for the system dynamics. We construct a quantum channel generating the non-commutative graph as a linear envelope of the POVM. The idea is based on applying a generalized version of a quantum channel using the apparatus of von Neumann algebras. The results are analyzes for a non-commutative graph generated by a qubit interacting with quantum oscillator. For this model the quantum anticlique which determines the error correcting subspace has an explicit expression.}
    \end{abstract}


\maketitle

\section{Introduction}

One of the important notions in quantum information theory is the notion of a non-commutative operator graph, which is an operator space containing the identity operator and closed under operator conjugation. For each completely positive trace-preserving map (i.e., a quantum channel) there is a unique operator graph determining the ability to transmit information with zero error via the channel. This graph allows to define the Knill-Laflamme sufficient condition for the subspace to be a quantum error-correction code. A natural opposite task is to find quantum channel corresponding to the given graph~\cite{amo, AM}. All the graphs are known to be linearly generated by positive operator valued measures (POVMs) and, vice versa, for each graph these exists POVM which generates this graph, so that the task can be posed for POVMs. A solution to this problem can be found using Naimark dilatation~\cite{AM}. 

Non-commutative operator graphs for various infinite-dimensional quantum systems were studied in~\cite{AMP2020,AMP2021}. In this paper, we study error correction for a model of an infinite-dimensional quantum system consisting of a qubit interacting with quantum oscillator~\cite{Gazeau1999}. The dynamics of the composite system is governed by Schr\"odinger equation which entangles initially separable quantum states. The dynamics generates POVM for the system. Quantum anticlique is the projector onto error correcting subspace. We construct a generalized quantum channel, acting between preduals of two von Neumann algebras, which determines the graph corresponding to the given POVM with an operator-valued density. Our construction is close to the similar finite-dimensional result presented in~\cite{SS}. The techniques are based upon~\cite{BL2006}. The results are analyzed for the graph corresponding to the error correction model of a qubit interacting with quantum oscillator. 

\section{Generalized quantum channels generated by POVMs}

We use some basic notions from the theory of von Neumann algebras ($W^*$-algebras in other terminology) \cite {Sakai}. 

Denote $B(\calH)$ and $T(\calH)$ the algebra of all bounded operators and the space of nuclear operators in a separable Hilbert space $\calH$ respectively, the notation $\Vert\cdot \Vert$ designates the operator norm. The subalgebra ${\mathcal M}\subset B(\calH)$ is said to be the von Neumann algebra if the second commutant satisfies ${\mathcal M}''={\mathcal M}$. Given a von Neumann algebra $\mathcal M$, there exists the predual Banach space ${\mathcal M}_*$ such that $({\mathcal M}_*)^*={\mathcal M}$ due to the Sakai theorem. The corresponding duality is denoted by $\braket {\rho ,x},\ \rho \in {\mathcal M}_*,\ x\in {\mathcal M}$. The functionals on $\mathcal M$ determined by elements of ${\mathcal M}_*$ are said to be normal. A normal positive functional $\rho $ with the property $\braket {\rho ,\mathbb I}=1$ is called a state.

The quantum channels can be considered as mappings  $\Phi:B(\calH_1)\rightarrow B(\calH_2)$ which can be represented in the Kraus form
$$
\Phi(\rho) = \sum\limits_{k} A_k\rho A_k^{*},\qquad A_{k}: \calH_1\rightarrow \calH_2,
$$ 
where $\sum\limits_k A_k A_k^{*}=\mathbb I.$ The non-commutative operator graph corresponding to the channel is 
$$\mathcal{V}=span\left\lbrace A_k^{*}A_j\right\rbrace.$$
The subspace $K\subset \calH_1$ is a quantum error-correcting code if the orthogonal projection $P_K$ satisfies the Knill-Laflamme condition $\dim P_K\mathcal{V}P_K=1$. Such projection $P_K$ is called quantum anticlique.

Suppose that ${\mathcal M}^{(1)}\subset B(\calH_1)$ and ${\mathcal M}^{(2)}\subset B(\calH_2)$ are two von Neumann algebras acting in the Hilbert spaces $\calH_1$ and $\calH_2$. Denote $\braket {\cdot ,\cdot }_{1,2}$ the corresponding dualities.
Given a linear map $\Phi :{\mathcal M}_*^{(1)}\to {\mathcal M}_*^{(2)}$, one can define the conjugate map $\Phi ^*:{\mathcal M}^{(2)}\to {\mathcal M}^{(1)}$ by the rule
$$
\braket {\rho ,\Phi ^*(x)}_1=\braket {\Phi (\rho ),x}_2,\qquad \rho \in {\mathcal M}_*^{(1)},\ x\in {\mathcal M}^{(2)}.
$$
Following~\cite{BL2006}, the map $\Phi $ is said to be a generalized quantum channel if $\Phi ^*$ is unital and completely positive.

Let $(\Omega, \cal B, \nu )$ be a measurable space with the $\sigma$-finite measure $\nu$. Then ${\mathcal M}=B(\calH)\otimes L^{\infty }(\Omega )$ is a $W^*$-algebra of operators acting in the Hilbert space $\tilde{\mathcal H}=\calH\otimes L^2(\Omega )$ and having the predual space ${\mathcal M}_*=T(\calH)\otimes L^1(\Omega )$. Also note that ${\mathcal M}$ can be viewed as $L^{\infty }(\Omega\rightarrow B(\calH))$, the space of $\nu$-essentially bounded $B(\calH)$-valued functions. Put ${\mathcal M}^{(1)}=B(\calH)$ and ${\mathcal M}^{(2)}=B(\calH)\otimes L^{\infty }(\Omega )$ such that ${\mathcal M}^{(1)}\subset B(\calH)$
and ${\mathcal M}^{(2)}\subset B(\calH\otimes L^2(\Omega ))$. Then, the quantum channel
$$
\Phi :{\mathcal M}_*^{(1)}\to {\mathcal M}_*^{(2)}
$$
is characterized by the property
\begin{equation}\label{1}
	\braket {\Phi (\rho ),I_{\calH\otimes L^2(\Omega )}}_2=\braket {\rho ,I_\calH}_1=1
\end{equation}
for all states $\rho \in {\mathcal M}_*^{(1)}$. Thus, $\Phi (\rho )\in {\mathcal M}^{(2)}_*$ is also a state. Since
$\Phi (\rho)$ is a function $f_{\rho }$ on the space $\Omega $ taking values in $T(\calH)$ the equality (\ref {1}) can be rewritten in the form
$$
\int\limits_{\Omega }\Tr(f_{\rho }(\omega ))\nu (d\omega)=1.
$$

Let $M$ be a positive operator-valued measure on $(\Omega, \cal B, \nu )$ with values in the set of positive operators $B(\calH)_+$.

\begin{theorem}\label{theor1} Suppose that there is an operator valued density $P(\omega),\ \omega \in \Omega $ of $M$ with respect to the measure $\nu $ such that
	$$
	M(d\omega )=P(\omega )\nu (d\omega ) .
	$$
	Then, the formula
	$$
	\Tr(\rho \Phi ^*(x\otimes f))=\int \limits _{\Omega }f(\omega )\Tr([P(\omega )]^{1/2}\rho [P(\omega )]^{1/2}x)\nu (d\omega )
	$$
	determines a unital normal completely positive map $\Phi ^*:{\mathcal M}^{(2)}\to {\mathcal M}^{(1)}$.
\end{theorem}

{\bf Proof}
	
	Let $\Omega =\cup _{k=1}^NB_k$ be the partitioning of $\Omega $ into a sum of disjoint $B_k\in {\mathfrak B}$ and a simple function $f_0{\big|}_{B_K}=a_k$. Define a unital normal completely positive map $\Phi ^*_N:{\mathcal M}^{(2)}\to {\mathcal M}^{(1)}$ as follows
	\begin{equation}\label{N}
		\Phi ^*_N(x\otimes f)=\sum \limits _k a_k\int \limits _{B_k}[P(\omega )]^{1/2}x[P(\omega )]^{1/2}\nu (d\omega ).
	\end{equation}
	So,
	$$
	\Tr(\rho \Phi ^*(x\otimes f_0))=\sum \limits _k a_k\int \limits _{B_k}\Tr([P(\omega )]^{1/2}\rho [P(\omega )]^{1/2}x)\nu (d\omega ),
	$$
	By this, the positivity of  all operators under the trace results in
	$$
	|\Tr(\rho \Phi ^*(x\otimes f_0))|\le \Vert f_{0}\Vert_{L^{\infty}}\int \limits _{\Omega}\Vert\rho\Vert\cdot\Vert x \Vert \cdot\Tr(P(\omega ))\nu (d\omega )=\Vert f_{0}\Vert_{L^{\infty}}\cdot\Vert\rho\Vert \cdot\Vert x\Vert,
	$$
	approaching $f\in L^{\infty}\left( \Omega\right) $ by simple functions not disrupt that inequality. So $\Phi ^*$ is a limit of (\ref {N}) in the weak* topology.

$\Box$

The  POVM $M$ generates some non-commutative graph ${\mathcal V}=\overline {span}(M(B),\ B\in {\mathfrak B})$, where $\mathfrak B$ is the $\sigma $-algebra of measurable subsets $B\subset \Omega $. Let us define a unital completely positive map $\hat \Psi ^*:L^{\infty }(\Omega )\to {\mathcal M}^{(1)}$ by the formula
\begin{equation}\label{kanal}
	\hat \Psi ^*(f)=\Phi ^*({\mathbb I}\otimes f),\qquad f\in L^{\infty }(\Omega ).
\end{equation}

\begin{theorem}\label{theor2}
The channel $\Psi $ complementary to $\hat\Psi$ defined by Eq.~(\ref{kanal}) determines the graph $\mathcal V$. 
\end{theorem}

{\bf Proof} The action $\hat \Psi ^*:L^{\infty }(\Omega )\to {\mathcal M}^{(1)}$ can be represented as follows
	$$
	\hat \Psi ^*(f)=\int \limits _{\Omega }f(\omega )P(\omega )\nu (d\omega )
	$$
	It suffices to show that ${\mathcal V}=\hat \Phi ^*(L^{\infty }(\Omega ))$ \cite{SS}. The result immediately follows from the equality
	$$
	\hat \Phi ^*(\chi _{B})=M(B),
	$$
	where $\chi _B\in L^{\infty }(\Omega )$ is the indicator function of the measurable set $B\in {\mathfrak B}$.
	
$\Box $

\section{A qubit interacting with quantum oscillator}

We consider a qubit interacting with quantum oscillator within the rotating wave approximation. This model is known to have an explicit description of the eigenstates and eigenvalues which completely define the model~\cite{Gazeau1999}. Let $\calH_f$ be the Hilbert space with the basis $\{\ket{k}, k\in \mathbb{N}_0\}$ (the quantum oscillator Hilbert space) and $\calH_s$ the two-dimensional Hilbert space with the basis $\{\ket{g},\ket{e}\}$ (qubit space). The Hilbert space of the composite system is $\calH=\calH_f \otimes \calH_s$. The Hamiltonian is
\begin{equation}\label{HAM}
	{\bf H}=\omega_f a^{+}a^{-}+\frac{\omega_s}{2} \sigma_z + \frac{\kappa}{2}(\sigma^{-} a^{+}+\sigma^{+}a^{-}),
\end{equation}
Here $\omega_s, \omega_f \in \mathbb{R_+}$ are the frequencies of the qubit and the quantum oscillator, respectively, $\kappa\ge 0$ is the coupling constant, $\sigma_z$ is the Pauli matrix, $\sigma^{+},\sigma^{-}$ are the rising and lowering operators of the qubit and the $a^{+},a^{-}$ are the creation and annihilation operators of the oscillator. The detuning parameter is $\Delta=\omega_f-\omega_s$. For the non-resonant case $\Delta\neq 0$, the eigenstates of the Hamiltonian are
\begin{eqnarray*}\label{dressedstates}
	&& \ket{0,g},\\
	&&\ket{n,+}=\cos \left(\frac{\theta_n}{2}\right)\ket{n-1,e}+\sin \left(\frac{\theta_n}{2}\right)\ket{n,g},\\
	&&\ket{n,-}=\sin \left(\frac{\theta_n}{2}\right)\ket{n-1,e}-\cos \left(\frac{\theta_n}{2}\right)\ket{n,g},
\end{eqnarray*}
where $\theta_n=\tan^{-1}(\kappa\sqrt{n}/\Delta)$ and $n\in \mathbb{N}$. For the resonant case $\Delta = 0$ the eigenstates are
\begin{eqnarray*}\label{dressedstates}
	&& \ket{0,g},\\
	&&\ket{n,+}=\ket{n-1,e}+\ket{n,g},\\
	&&\ket{n,-}=\ket{n,g}-\ket{n-1,e}.
\end{eqnarray*}
In both cases the corresponding eigenenergies are
\begin{eqnarray*}
	E_{0,g}&=& \frac{\omega_f+\Delta}{2} \\
	E_{n,\pm}&=&\omega_f\left(n-\frac{1}{2}\right)\pm \frac{1}{2}\sqrt{\Delta^2 + \kappa^2 n},\quad  n\in \mathbb{N}.
\end{eqnarray*}  

Our construction can be applied to this model of a qubit interacting with quantum oscillator. Let us split the Hilbert space $\calH$ into three parts~\cite{AMP2021},
$$
\calH=\calH_1\oplus \calH_2\oplus \calH_3,
$$
The partition is determined by the parameter $K_0\ge \max\{3,M_0\}$, where $M_0$ is the minimal natural solution of the inequality
\begin{equation}\label{K0}
	\left(\sqrt{\Delta^2+\kappa^2(M_0+1)}+\sqrt{\Delta^2+\kappa^2M_0}\right)^{-1}<\frac{2\omega_f}{\kappa^2}.
\end{equation}
The subspaces $\calH_1$ and $\calH_2$ are the infinite-dimensional subspaces corresponding to two strictly increasing sequences of eigenvalues $J_k=E_{k+1,+}$, $k\in \mathbb{N}_0$ and $S_{k+K_0}=E_{k+K_0,-}$, $k\in \mathbb{N}_0$. The subspaces are defined as follows
\begin{eqnarray*}
	\calH_1 &=&span\{\ket{n,+},\ n\in \mathbb{N}\},\\
	\calH_2 &=&span\{\ket{n,-},\ n\ge K_0\},\\
	\calH_3 &=&span\{\ket{g,0}\}\cup\{\ket{n,-},\ 1\le n < K_0\}.
\end{eqnarray*}
The sequences $J_k$ and $S_{k+K_0}$ allow to define Gauzeau-Klauder coherent states in $\calH_1$ and $\calH_2$ 
\begin{eqnarray*}
\ket {J,x,y}&=&\frac {1}{N_1(x)}\sum \limits _{k=0}^{+\infty }\frac {x^{k/2}e^{-iJ_ky}}{\sqrt {c_k^{(1)}}}\ket {k+1,\ +}\\
\ket {S,x,y}&=&\frac {1}{N_2(x)}\sum \limits _{k=0}^{\infty }\frac {x^{k/2}e^{-iS_{k+K_0}y}}{\sqrt {c_k^{(2)}}}\ket {k+K_0,\ -}.
\end{eqnarray*}
Here sequences $c_k^{(2)}$ are the positive converging weights, $N_1(x)$ and $N_2(x)$ are the normalization factors.

Let $\Omega =\mathbb {R}^2\oplus \mathbb {R}^2\oplus \{pt\}$, where $\{pt\}$ is the set containing only one point.
Define the POVM over $\Omega $ as follows
$$
M=M_1\oplus M_2\oplus M_3,
$$
where
\begin{eqnarray*}
M_1(dxd\mu (y))&=&\ket {J,x,y}\bra {J,x,y}\tau _1(x)dxd\mu (y),\\
M_2(dxd\mu (y))&=&\ket {S,x,y}\bra {S,x,y}\tau _2(x)dxd\mu (y),
\end{eqnarray*}
and
$$
M_3(\emptyset)=0,\qquad M_3(\{pt\})=P_3,
$$
Here $P_3$ is the projection on $\calH_3$ and measures $\tau _1(x)$, $\tau _1(x)$ are determined by the Gauzeau-Klauder construction~\cite{Gazeau1999}. The POVM $M$ is generated by orbits of the unitary group ${\bf U}_t=e^{-it\bf H}$ with the Hamiltonian (\ref {HAM}) and satisfies the conditions of Theorems~\ref{theor1} and~\ref{theor2}. The corresponding graph has the quantum anticlique $P_3$.

\section{Conclusion}
Based on the theory of non-commutative operator graphs, we analyze the error correction model for a qubit interacting with quantum oscillator. The dynamics of the composite system is governed by Schr\"odinger equation which generates POVM. We describe the method of how to define the quantum channel which corresponds to a non-commutative operator graph generated by the POVM. We analyze this construction for the model of a qubit interacting with quantum oscillator and provide an explicit expression for the quantum anticlique which determines for this model the error correcting subspace.

\section*{Acknowledgments}
The work is performed in Steklov Mathematical Institute of Russian Academy of Sciences within the project of the Russian Science Foundation 17-11-01388.

\end{document}